\documentclass[aps,prb,twocolumn,showpacs]{revtex4}
\usepackage{epsfig}
\usepackage{amsmath}
\usepackage{times}

\begin{document}

\title{Size-dependent thermal conductivity of nanoscale semiconducting systems}

\author{L. H. Liang$^1$ and Baowen Li$^{1,2,3}$\footnote{Correspondence should be addressed
to: phylibw@nus.edu.sg}}
 \affiliation {$^1$Department of Physics and
Center for Computational Science and Engineering, National
University of Singapore, Singapore 117542,
 Republic of Singapore\\
$^2$ Laboratory of Modern Acoustics and Institute of Acoustics,
Nanjing University, 210093, PR China
\\
 $^3$ NUS Graduate School for
Integrative Sciences and Engineering, Singapore 117597, Republic
of Singapore}

\date{6 March 2006}

\begin{abstract}
We study the size dependence of thermal conductivity in nanoscale
semiconducting systems.  An analytical formula including the
surface scattering and the size confinement effects of phonon
transport is derived. The theoretical formula gives good
agreements with the existing experimental data for Si and GaAs
nanowires and thin films.
\end{abstract}

 \maketitle

It is of great importance from both fundamental and application
point of views to understand  heat conduction in nanoscale
materials. On one hand, many fundamental questions remain open,
such as whether the thermodynamical laws still apply to the
nanoscale materials which are characterized by finite number of
atoms/molecules, while the thermodynamics is established on the
fact that the system contains infinite large number of molecules.
On the other hand, a continuous miniaturization and increase of
running speed of semiconductor devices will give rise to redundant
heat that will deteriorate the devices'  efficiency.\cite{Lieber}
How to dissipate the heat from these devices becomes a crucial
issue in heat management.

Generally, two approaches are used in studying heat conduction in
such low-dimensional systems of finite size. At first, in order to
understand the physical mechanism of heat conduction in such
systems, many one-dimensional (1D) lattice models with and without
on-site (pinning) potential are used such as the Fermi-Pasta-Ulam
(FPU) model \cite{FPU} without on-site potential, and the Frenkel
Kontoroval (FK) model and the $\phi^4$ model with on-site
potential.\cite{F4FK} More recently, a polymer model with
transverse motion has been introduced and studied both
analytically and numerically.\cite{WangLi04} All these models
nicely demonstrate the role of anharmonicity and the role of
on-site potential in 1D heat conduction. They are useful in
helping us understand the fundamental law of heat conduction, the
Fourier law. In the system without on-site potential, a
size-dependent thermal conductivity has been observed,\cite{FPU}
which is attributed to a superdiffusive motion of
phonons.\cite{LW03} However, such models are too simple to be used
for modelling heat conduction in realistic nanostructures, because
the heat conduction in such models is restricted to 1D or
quasi-1D, the scattering of phonons from surfaces is completely
neglected.

Another popular approach employs the Boltzmann transport equation,
which can explain the bulk transport phenomena well. However, as
we know that for nanostructures, both the effects of phonon
confinement and non-equilibrium phonon distribution due to
boundary scattering are important. Indeed, recent theoretical and
experimental investigation demonstrates a size-dependent behavior
of lattice vibration of nanostructures.\cite{SunCQ05} Experimental
measurements also show that thermal conductivity of semiconducting
nanowires and thin films is smaller than the corresponding bulk
value.\cite{Fon02,Li03,Cahill03,Liu04} However, a general
satisfactory theory to give a prediction agreed with existing
experimental results at room temperature is not available yet.

In this paper, we propose a phenomenological theory for the size
dependence of thermal conductivity by taking into account the
intrinsic size effect of phonon velocity, mean free path and the
surface scattering effect. An explicit analytic expression is
obtained, in which all parameters have clear physical meaning, and
good agreements with the existing experiments are found. The
different roles of two basic effects in the different size range
are illustrated.

We start with the well-known  kinetic formula of thermal
conductivity $\kappa$ for bulk dielectric materials,\cite{Zimann}
\begin{equation}
\kappa = \frac{1}{3}c \upsilon l \label{eq:kappa0},
\end{equation}
where $c$ is the specific heat, $\upsilon$ the average phonon
velocity and $l$ the mean free path (MFP). We assume that phonons
predominate the thermal conduction, which is true for
semiconductors or insulators discussed here. For thermal
conductivity of nanomaterials, we first consider the size
dependence of $\upsilon$ and $l$. The specific heat is assumed to
be constant and the room temperature is applied in this paper. The
average phonon velocity is proportional to the characteristic
Debye temperature of crystals,\cite{Debyesound,Regel95}
\begin{equation}
\Theta \propto {\frac{2h}{\pi k_B}} \left(\frac{3N_A}{4 \pi
V}\right) ^{1/3} \upsilon \label{eq:Debyesound}
\end{equation}
with the Planck constant $h$, the Boltzmann constant $k_B$, the
Avogadro constant $N_A$, and the molar volume $V$. The system
considered is assumed to be isotropic.

In the following discussion, we suppose  Eq. (\ref{eq:Debyesound})
is valid for the corresponding nanoscale crystals.  Let $L$ be the
size of nanostructures, such as the diameter of nanowires or the
thickness of thin films, the $L\to \infty$ limit is denoted by
using the subscript $b$ which means the corresponding bulk limit,
the size dependence of the phonon velocity is equal to that of the
Debye temperature, we thus have, $\upsilon_L/\upsilon_b =
\Theta_L/\Theta_b$.

From the Lindemann's proposition, we may get the relationship
between the melting point and the Debye temperature of crystals.
In 1910, Lindemann proposed a melting criterion, known to be valid
for small particles,\cite{Frenkel93} stating that a crystal will
melt when the root mean square displacement (MSD) $\sigma$ of
atoms in the crystal exceeds a certain fraction of the interatomic
distance.\cite{Lindemann} Combining with the Einstein specific
heat theory, the square of the characteristic temperature is
proportional to the melting point $T^m$ of crystals, and the
modern form of this relation for the Debye temperature is
\cite{Dash99,Zimann}
\begin{equation}
\Theta = \mbox{const.}\left(\frac{T^m}{M V^{2/3}}\right)^{1/2}
\label{eq:Debyemelt}
\end{equation}
with the molecular mass $M$. According to the same relation for
nanocrystals,
\begin{equation}
\Theta^2_L/ \Theta^2_b = T^m_L/T^m_b. \label{eq:ThetaT}
\end{equation}

Combining the Lindemann melting formula and the Debye expression
for thermal conductivity, eliminating the Debye temperature and
the sound velocity, it can be deduced that the MFP is proportional
to the melting point at a given absolute temperature $T$, $l
\propto 20{T^m}a/(\gamma^{2}T) $ with the lattice constant $a$ and
the Gruneisen constant $\gamma$.\cite{MFP,Zimann} Similarly, the
same relation is assumed for nanocrystals, we have
\begin{equation}
l_L/l_b = T^m_L/T^m_b. \label{eq:MFPT}
\end{equation}

From Eqs. (\ref{eq:kappa0}), (\ref{eq:Debyesound}),
(\ref{eq:ThetaT}) and (\ref{eq:MFPT}), we obtain an expression for
the size-dependent thermal conductivity of nanosemiconductors
${[T^m_L/T^m_b]}^{3/2}$. Moreover, based on the fact of the
size-dependent atomic thermal vibrations, the relations among the
atomic MSD $\sigma^2$, the Debye temperature and the melting
point, the size-dependent melting temperature function of
nanocrystals has been modelled and validated as follows
\cite{JiangQ}
\begin{equation}
 T^m_L/T^m_b = \left(\sigma_b
/\sigma_L\right)^2 = \exp \left( {\frac{ - (\alpha - 1)}{L / L_0 -
1}} \right), \label{eq:TL}
\end{equation}
where $\alpha(\equiv \sigma_s^2/\sigma_i^2)$  is a material
constant with $\sigma_s^2$ and $\sigma_i^2$ corresponding to the
MSD of surface atoms of a crystal and that of atoms within the
crystal, respectively. For the free-standing nanocrystals, $\alpha
= 2S_v/(3R) + 1 > 1$ with the bulk vibrational entropy $S_v$ of
melting and the ideal gas constant $R$ \cite{JiangQ} based on the
Mott's expression for the vibrational entropy and its relation
with the melting point.\cite{Mott,Regel95} $L_0$ is a critical
size at which almost all atoms of a crystal are located on its
surface, $L_0 = 2(3-d)w$ with the atomic/molecular diameter $w$
and the dimension $d = 0, 1, 2$ for nanoparticles, nanowires and
thin films, respectively.\cite{JiangQ} Note that the two basic
assumptions in the model of the size-dependent atomic thermal
vibrations of nanocrystals are: (1) although ${\sigma_s}^2$ and
${\sigma_i}^2$ are considered to be size-dependent (the phonon
softening is considered to occur not only on the surface but also
in the interior for small size crystals), $\alpha$, the ratio
between them, is taken approximately as a size-independent value;
(2) the cooperative coupling between the surface region and the
interior region is considered phenomenologically by taking the
variation of ${\sigma}^2$ to be dependent on the value of itself,
${\sigma}^2$ is the average MSD over the crystal with the
respective weight of ${\sigma_s}^2$ and
${\sigma_i}^2$.\cite{JiangQ} With the above consideration, a
change in ${\sigma}^2$ can be given by
${\sigma}^2(x+dx)-{\sigma}^2(x) = (\alpha -1){\sigma}^2(x)dx$,
where the surface-volume ratio $x=L_0/(L-L_0)$. By integrating the
above equation, the size-dependent MSD was obtained as shown in
Eq. (\ref{eq:TL}). According to ${\sigma}^2 \propto T/{\Theta^2}$
in the high temperature approximation $(T > \Theta/2)$
\cite{Childress91,Dash99} and Eq. (\ref{eq:ThetaT}), ${\sigma
^2_b}/{\sigma ^2_L} = {\Theta^2_L}/{\Theta^2_b} = {T^m_L
}/{T^m_b}$. Equation (\ref{eq:TL}) indicates that the melting
point and thus the thermal conductivity decreases with reducing
size $L$ of nanocrystals, it is valid for crystals of $L \geq
2L_0$ with $2L_0$ representing a characteristic length scale for
the crystallinity.\cite{JiangQ}

The above discussion on the thermal conductivity of
nanostructures, based on the effective bulk formula, includes the
intrinsic size effect of the phonon velocity and the MFP. The
phonon-phonon interaction increases with size reduction due to the
confinement, which causes the increase of thermal resistance and
the decrease of heat conduction. On the other hand, as the size
decreases and the surface-volume ratio increases, the large
surface/interface scattering, corresponding to certain boundary
conditions in the linearized Boltzmann equation, has great
influence on the transport. Considering the nonequilibrium phonon
distribution due to boundary scattering, the effect of the surface
roughness with the boundary scattering shows an exponential
suppression in the distribution and the
conduction.\cite{Zimann,Ziambaras} Therefore, a term
$p\exp(-l_0/L)$ is added to correct the bulk formula, where $p$ is
a factor reflecting the surface roughness, $l_0$ is the phonon MFP
in the Debye model at room temperature and assumed to be a
constant here since we have considered the size effect in the
above discussion. $l_0/L$ corresponds to the Knudsen number of the
phonon Knudsen flow induced by the interface
scattering.\cite{Ziambaras} Finally, the size-dependent thermal
conductivity is obtained as
\begin{equation}
 \frac{\kappa_L}{\kappa_b}= p\exp{\left(-\frac{l_0}{L}\right)}\left[{\exp
 \left({\frac{ -(\alpha-1)}{L / L_0 -
1}} \right)}\right]^{3/2}.
\label{eq:KappaL}
\end{equation}

\begin{table}
\caption{Parameters used in Eq. (\ref{eq:KappaL}) for predictions
in this paper. $\alpha = 2S_v/(3R) + 1$, $ S_v = S_m -R$ (Ref. 22)
and $S_m = H_m/T^m$ with the bulk melting entropy
$S_m$, enthalpy $H_m$ and temperature $T^m$. $L_0 = 4w, 2w$ for
nanowires and thin films, respectively.}

\begin{tabular}
{|c|c|c|c|c|} \hline  &  $H_m$(KJmol$^{-1}$) &  $T_m$(K) &
 $w$ (nm) &  $l_0$ (nm)  \\
\hline  Si & 50.55 \cite{Table}& 1685 \cite{Table}&
0.3368 \cite{King} &  41 \cite{Zou01} \\
\hline  GaAs &  120 \cite{Nakamura}&  1511 \cite{Table,Nakamura}&
0.245 \cite{Ohtake} & 5.8 \cite{Sze}  \\
\hline
\end{tabular}

\label{tab1}
\end{table}

\begin{figure}
\includegraphics
[width=\columnwidth]{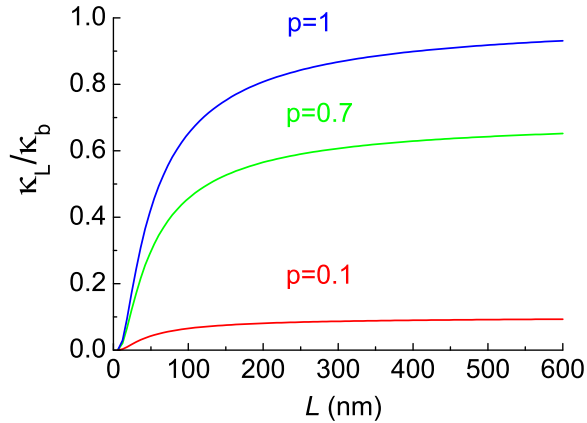} \vspace{-1cm}
\caption{Size-dependent dimensionless thermal conductivity for Si
films in terms of Eq. (\ref{eq:KappaL}), $\kappa_L/\kappa_b$
versus system size $L$ for $p=0.1$, 0.7 and 1, respectively. The
other parameters in Eq. (\ref{eq:KappaL}) are given in Table I.}
\label{fig:Theory}
\end{figure}

This is the central result of this paper. It gives a quantitative
prediction for the size-dependent thermal conductivity. $0 < p
\leq 1$. The larger value of $p$ corresponds to the smaller
roughness, i.e. the smoother surface, thus the more probability of
specular scattering, vice versa, the smaller $p$ corresponds to
the more probability of diffusive scattering. The smaller $L$
corresponds to the relatively larger surface roughness usually and
thus the smaller $p$. $p$ depends greatly on fabrication precision
of nanostructures, its physical meaning will be discussed further
later on. When system size deceases, the value of $\exp(-l_0/L)$
decreases, which reflects the increase of the interface scattering
and thus the weakened conduction. From the point of view of the
bulk approach, there are two asymptotic limits to be satisfied by
Eq. (\ref{eq:KappaL}): $L \rightarrow \infty$ and $p \rightarrow
1, \kappa_L \rightarrow \kappa_b$; $L \rightarrow L_0$ or $p
\rightarrow 0, \kappa_L \rightarrow 0$. In Fig. \ref{fig:Theory},
we show $\kappa_L/\kappa_b$ versus system size $L$ for different
roughness parameter $p$. The thermal conductivity spans a wide
range, in which our prediction at $p = 0.7$ for $L < 100$ nm is
consistent with the previous theoretical
prediction.\cite{Ziambaras} The intrinsic size effect decreases
the conductivity obviously when the size is smaller than about 100
nm. The surface effect related to $p$ can explain why the
conductivity is low even in the large size in some case.

\begin{figure}
\includegraphics
[width=\columnwidth]{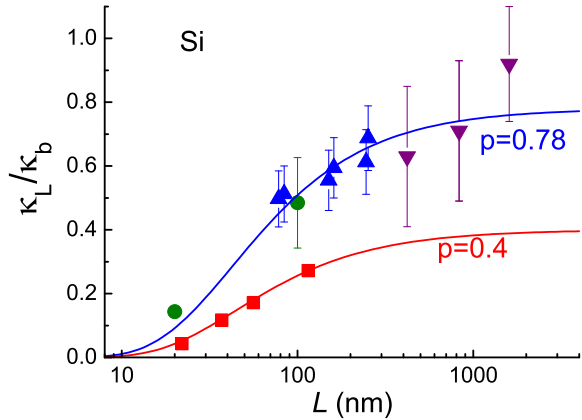}
 \caption{Size-dependent
thermal conductivity of Si nanowires and thin films at 300 K. The
symbols are experimental results, the up-triangles, circles and
down-triangles for thin films (in-plane) cited from Refs.9,10 and
29,
respectively, the squares for nanowires (along the
axis).\cite{Li03} The curves are predictions from Eq.
(\ref{eq:KappaL}), the upper one for thin films with $p=0.78$, the
lower one for nanowires with $p=0.4$.} \label{fig:SiKappa}
\end{figure}

Figure \ref{fig:SiKappa} shows the size-dependent thermal
conductivity of Si nanowires and thin films from the experiments
and the comparison with our theoretical predictions from Eq.
(\ref{eq:KappaL}) with suitable parameters $p$. The predictions
are in good agreements with the experimental results. For
carefully prepared Si films with thickness of 70-250
nm,\cite{Cahill03} $p = 0.78$ corresponds with the experimental
results, this large value of $p$ implies small diffusive
scattering contribution due to the smooth surface related with the
careful preparation of the nanostructures. For Si nanowires with
diameter of 20-100 nm,\cite{Li03} $p = 0.4$, reflecting the larger
surface roughness of the nanowires in the sample fabrication.

Now we turn to the underlying physical meaning of the roughness
parameter $p$. For the Si films with the thickness $L = 100$ nm,
if the surface roughness $\eta$, the mean root square deviation of
height of the surface from the reference plane, is 2.2 nm, which
may be the case according to the precise fabrication technology of
monocrystalline Si layers with thickness dispersion smaller than 4
nm,\cite{Bruel} $p = 0.78$ can be obtained by a relation $p =
1-10\eta/L$; according to this relation, if $p = 0.4$, $\eta$ is
1.32 nm for $L = 22$ nm, which agrees with the experimental
observation for the Si nanowires well,\cite{Li03} thus, this
fitted formula is valid and can be used to determine $p$ by
measuring or estimating $\eta$ of samples. The larger $\eta$ for a
given $L$ corresponds to the smaller $p$, i.e., the rougher
surface corresponds to the higher probability of diffusive
scattering. Therefore, carefully fabricated structures with
smoother surface will bring forth a larger conductivity under the
same other conditions.

Figure \ref{fig:SiKappa} also shows that $p = 0.78$ is applicable
for the films spanning the size range of 20-100 nm, 70-250 nm and
400-1600 nm, since the decreased $\eta$ can be achieved with
decreasing $L$. Note that the surface roughness may be different
for samples of different size even in the same experiment, but it
can be controlled to be fewer than tens of nanometers generally in
modern fabrication.\cite{Bruel}

\begin{figure}
\includegraphics
[width=\columnwidth]{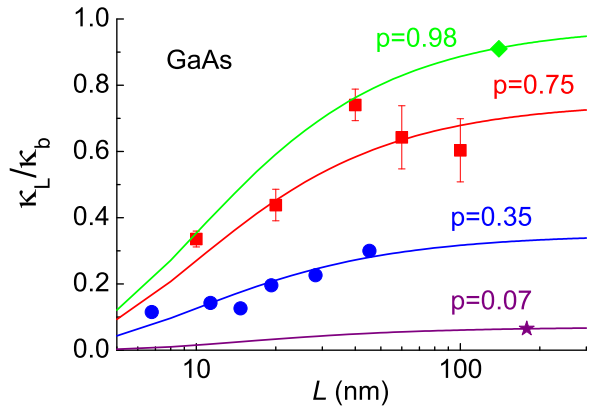} \vspace{-1cm}
\caption{Size-dependent thermal conductivity of GaAs nanowires and
thin films. The star is experimental result of nanowire with
diameter about 180 nm (corresponding to the rectangle
cross-section with the same area) at 40 K,\cite{Fon02} considering
the similar conductivity difference between the nanowire and the
bulk extrapolated at 300 K, the curve is prediction from Eq.
(\ref{eq:KappaL}) with $p=0.07$. The diamond is experimental
result of GaAs/AlAs superlattice with thickness 140 nm,\cite{Yu95}
assumed as the conductivity of a single GaAs film with the same
thickness considering the similar structures between GaAs and AlAs
(the following is the same), the curve for prediction with
$p=0.98$. The squares are experimental results of GaAs/AlAs
superlattices with thickness 10-100 nm (in-plane),\cite{Yao87} the
curve for prediction with $p=0.75$. The circles are experimental
results of GaAs/AlAs superlattices with thickness 5-45 nm
(cross-plane),\cite{Capinski} the curve for prediction with
$p=0.35$.} \label{fig:GaAsKappa}
\end{figure}

Figure \ref{fig:GaAsKappa} shows the thermal conductivity of GaAs
nanowires and thin films. The theoretical predictions with
corresponding $p$ also show agreements with the experimental
results from different groups. Similarly to Si nanowires, $p =
0.07$ is also small for GaAs nanowires,\cite{Fon02}  $\kappa_L <
0.1\kappa_b$. It seems that for nanowires, the thermal conduction
is much worse compared to that of the corresponding films, which
may be attributed to the restrained dimension, the surface
fabrication precision and thus the increased surface scattering.
For the superlattices with thickness of 140 nm,\cite{Yu95} $p =
0.98$, $\kappa_L \approx 0.9\kappa_b$ is very high. For the
superlattices of 10-100 nm,\cite{Yao87} $p = 0.75$ and $\kappa_L$
is lower. The smaller $L$ corresponds to the relatively larger
surface roughness and the smaller $p$, thus the lower
conductivity. For the superlattices of 5-45 nm,\cite{Capinski} $p
= 0.35$, $\kappa_L$ corresponds to the cross-plane conductivity
measured and is even lower. Note that this low conductivity
results also from the scattering contributions of the
multi-interface corresponding to the increased interface roughness
$\eta$, which has the weaker effect on the in-plane conductivity.
When the size decreases to about 5 nm, the structure and energy
state of these nanosolids may be different from that of the
corresponding crystals, the small size and large surface effect
may not be sufficient to describe the physical properties of such
small size structures and the quantum effect may be dominant, the
present model is not suitable for that case. Some molecular
dynamic calculation shows the minimum with the size in the
cross-plane thermal conductivity for the thinner superlattices
without roughness, which is also theoretically attributed to a
crossover between the particle transport and the wave
transport.\cite{Maris}

In summary, we have derived a quantitative formula for the
size-dependent thermal conductivity of nanoscale semiconducting
systems by taking into account the intrinsic size effect and by
correcting the bulk expression. The formula relates the thermal
conductivity with the system size by the surface roughness
parameter, the Knudsen number, the crystallinity length scale and
the atomic vibration parameter.  The model reveals the respective
roles in the different size range of two basic effects on
nanoscale thermal transport, i.e., the intrinsic size effect is
predominant at the size of about 5-100 nm, and the surface
scattering effect is dominant at the larger size.  The theory
agrees well with the existing experimental results for Si and GaAs
nanowires and thin films.

This work is supported in part by grants from the Faculty Research
Grant of NUS and the DSTA Singapore under agreement DSTA
POD0001821.

\end{document}